\documentclass[aps,prl,showpacs,preprint]{revtex4}
\usepackage{graphicx}
\usepackage{amsmath}
\newcommand{\be}{\begin{equation}}
\newcommand{\ee}{\end{equation}}
\newcommand{\bea}{\begin{eqnarray}}
\newcommand{\eea}{\end{eqnarray}}
\newcommand{\bc}{\begin{center}}
\newcommand{\ec}{\end{center}}

\begin{document}

\preprint{SLAC-PUB-11598}

\title{Fully Coherent X-ray Pulses from a Regenerative Amplifier Free Electron Laser}

\author{Zhirong Huang and Ronald D. Ruth}
\affiliation{Stanford Linear Accelerator Center, Stanford
University, Stanford, CA 94309}

\date{January 30, 2006}

\begin{abstract}
We propose and analyze a novel regenerative amplifier free
electron laser (FEL) to produce fully coherent x-ray pulses. The
method makes use of narrow-bandwidth Bragg crystals to form an
x-ray feedback loop around a relatively short undulator.
Self-amplified spontaneous emission (SASE) from the leading
electron bunch in a bunch train is spectrally filtered by the
Bragg reflectors and is brought back to the beginning of the
undulator to interact repeatedly with subsequent bunches in the
bunch train. The FEL interaction with these short bunches not only
amplifies the radiation intensity but also broadens its spectrum,
allowing for effective transmission of the x-rays outside the
crystal bandwidth. The spectral brightness of these x-ray pulses
is about two to three orders of magnitude higher than that from a
single-pass SASE FEL.
\end{abstract}

\maketitle

An x-ray free electron laser (FEL) based on self-amplified
spontaneous emission (SASE) is an important first step towards a
hard x-ray laser and is expected to revolutionize the ultrafast
x-ray science (see, e.g., Refs.~\cite{LCLS-CDR,TESLA-XFEL}).
Despite its full transverse coherence, a SASE x-ray FEL starts up
from electron shot noise and is a chaotic light temporally. Two
schemes have been proposed to improve the temporal coherence of a
SASE FEL in a single pass configuration. A high-gain harmonic
generation (HGHG) FEL uses available seed lasers at ultraviolet
wavelengths and reaches for shorter wavelengths through cascaded
harmonic generation~\cite{yu02}. In this process, the ratio of
electron shot noise to the laser signal is amplified by at least
the square of the harmonic order and may limit its final
wavelength reach to the soft x-ray region~\cite{saldin02}. Another
approach uses a two-stage SASE FEL and a monochromator between the
stages~\cite{feldhaus}. The SASE FEL from the first undulator is
spectrally filtered by a monochromator and is then amplified to
saturation in the second undulator. This approach requires an
undulator system almost twice as long as a single-stage SASE FEL.

\begin{figure}
\centering
    \includegraphics[width=0.5\linewidth]{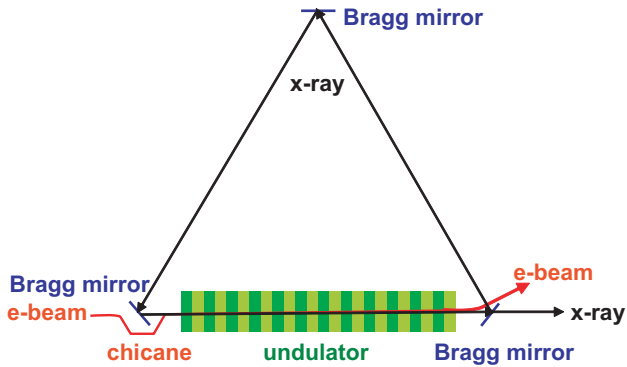}
    \caption{(Color) Schematic of an x-ray RAFEL using three Bragg crystals.}
    \label{fig:rafel}
    \vspace{-4mm}
\end{figure}

Another seeding scheme, a regenerative amplifier FEL (RAFEL), has
been demonstrated in the infrared wavelength region~\cite{nguyen}
and discussed in the ultraviolet wavelength region~\cite{faatz99}.
It consists of a small optical feedback and a high-gain FEL
undulator. In the hard x-ray region, perfect crystals may be used
in the Bragg reflection geometry for x-ray
feedback~\cite{colella,adams97} and have been demonstrated
experimentally for x-ray photon storage (see, e.g.,
Ref.~\cite{liss,shvydko}). In this paper, we propose and analyze a
novel x-ray RAFEL using narrow-bandwidth, high-reflectivity Bragg
mirrors. The basic schematic is shown in Fig.~\ref{fig:rafel}.
Three Bragg crystals are used to form a ring x-ray cavity around a
relatively short undulator. Alternative backscattering geometry
with a pair of crystals may also be used. SASE radiation from the
leading electron bunch in a bunch train is spectrally filtered by
the Bragg reflectors and is brought back to the beginning of the
undulator to interact with the second bunch. This process
continues bunch to bunch, yielding an exponentially growing laser
field in the x-ray cavity. The FEL interaction with these short
bunches not only amplifies the radiation intensity but also
broadens its spectrum. The downstream crystal transmits the part
of the radiation spectrum outside its bandwidth and feeds back the
filtered radiation to continue the amplification process. Compared
to a SASE x-ray FEL that typically requires more than 100~m of
undulator distance, this approach uses a significantly shorter
undulator but a small number of electron bunches to generate
multi-GW x-ray pulses with excellent temporal coherence. The
resulting spectral brightness of these x-ray pulses can be another
two to three orders of magnitude higher than the SASE FEL.

We first consider a one-dimensional (1-D) model of the
narrow-bandwidth RAFEL to describe its main characteristics such
as the temporal profile, the round-trip power gain and the maximum
extraction efficiency. At the beginning of the $n^{th}$ undulator
pass, the radiation field is represented by $E_n(t)$, where $t$ is
the arrival time relative to the longitudinal center of the
electron bunch. The radiation field at the exit of the undulator
is
    \be
    E_n^a (t) \approx E_n(t) g(t)+\delta E_n(t)\,,
    \label{eq:field}
    \ee
where $\delta E_n(t)$ is the SASE signal of the $n^{th}$ electron
bunch. When the radiation slippage length is much smaller than the
electron bunch length, we can assume the electric field gain
factor $g(t)$ is a function of the local beam current which can be
approximated by
    \be
    g(t)\approx g_0 \exp\left(-{t^2\over 2\sigma_\tau^2}\right)\,,
    \label{eq:gain}
    \ee
where $\sigma_\tau$ is the rms pulse duration of a Gaussian bunch
current. The more precise gain dependence on the current is used
in numerical simulations shown below.

The amplified signal is then spectrally filtered by the Bragg
mirrors and is fed back to the entrance of the undulator in the
$(n+1)^{th}$ pass, i.e.,
    \be
    E_{n+1}(t)=\int_{-\infty}^\infty {d\omega\over 2\pi}e^{-i\omega t}
    \int_{-\infty}^\infty dt' E_n^a(t') e^{i\omega t'}
    f(\omega-\omega_r)\,, \label{eq:feedback}
    \ee
where $f(u) =r\exp\left(- {u^2/ 4\sigma_m^2}\right)$ is a Gaussian
spectral filter function with the rms intensity bandwidth
$\sigma_m$, and $\omega_r$ is the central frequency of the filter
with the power reflectivity $\vert r\vert^2\le 1$. For a high-gain
amplifier after a few passes, the seed signal dominates over the
SASE, so that we can neglect the second term on the right side of
Eq.~(\ref{eq:field}). Integrating Eq.~(\ref{eq:feedback}) over the
frequency yields
    \be
    E_{n+1}(t) = \int_{-\infty}^\infty dt' {r\sigma_m\over \sqrt{\pi}}
    e^{-i\omega_r (t-t')} e^{-\sigma_m^2 (t-t')^2} g(t') E_n(t')
    \,.\label{eq:difference}
    \ee
Since there is no initial seed signal, $E_1(t)=0$, and
    \be
    E_2(t) = \int_{-\infty}^\infty dt' {r\sigma_m\over \sqrt{\pi}}
    e^{-i\omega_r (t-t')} e^{-\sigma_m^2 (t-t')^2} \delta E_1(t')
    \ee
is the spectrally filtered SASE from the first pass that seeds the
second pass.

For $n\gg 1$, we look for an exponentially growing solution
    \be
    E_n(t)=\Lambda^n A(t)e^{-i\omega_r t}\,.
    \ee
Eq.~(\ref{eq:difference}) is then transformed to an integral
equation:
    \be
    \Lambda A(t) = \int_{-\infty}^\infty dt' K(t,t') A(t')\,,
    \label{eq:mode}
    \ee
with the kernel
    \be
    K(t,t') ={r\sigma_m\over \sqrt{\pi}} e^{-\sigma_m^2 (t-t')^2}
    g(t')\,.
    \ee
Since both $r$ and $g(t')$ may be complex, $K(t,t')$ is in general
not a hermitian kernel.

We expect that a Gaussian fundamental mode will have the largest
gain $\vert \Lambda_0\vert$, i.e.,
    \be
    A_0(t) = \exp\left(-{t^2\over 4\sigma_{x0}^2}\right)\,.
    \label{eq:eigen}
    \ee
Here $\sigma_{x0}$ is the rms pulse duration of the returning
filtered radiation. Inserting Eq.~(\ref{eq:eigen}) into
Eq.~(\ref{eq:mode}), we obtain
    \begin{align}
    \Lambda_0 \exp\left(-{t^2\over 4\sigma_{x0}^2}\right) = g_0 r {2\sigma_m\sigma_{xa}\over
    \sqrt{1+4\sigma_m^2\sigma_{xa}^2}} \exp\left(-{\sigma_m^2 t^2\over
    1+4\sigma_m^2\sigma_{xa}^2}\right)\,,
    \label{eq:algebra}
    \end{align}
where $\sigma_{xa} = \sigma_{x0} \sigma_\tau/\sqrt{2\sigma_{x0}^2
+ \sigma_\tau^2}$ is the rms x-ray pulse duration at the undulator
end (see Eq.~(\ref{eq:all rad})). The self-consistent solution of
Eq.~(\ref{eq:algebra}) is
    \begin{align}
    & \sigma_{x0}^2 = {1+4\sigma_m^2\sigma_{xa}^2\over 4
    \sigma_m^2} ={\sqrt{1+8\sigma_m^2\sigma_\tau^2}+1\over 8\sigma_m^2}\,, \nonumber \\
    & \sigma_{xa}^2 = {\sqrt{1+8\sigma_m^2\sigma_\tau^2}-1\over 8\sigma_m^2}
    \,, \label{eq:length2}\\
    & \Lambda_0 =g_0 r {2\sigma_m\sigma_{xa}\over
    \sqrt{1+4\sigma_m^2\sigma_{xa}^2}} \,.\nonumber
    \end{align}
Thus, the round-trip power gain is
    \begin{align}
    G_{\text{eff}} \equiv & \vert\Lambda_0\vert^2 =G_0 R {4\sigma_m^2\sigma_{xa}^2\over 1+4\sigma_m^2\sigma_{xa}^2}
    \nonumber \\
    =& G_0 R{\sqrt{1+8\sigma_m^2\sigma_\tau^2}-1
    \over
    \sqrt{1+8\sigma_m^2\sigma_\tau^2}+1}
    \,. \label{eq:eff gain}
    \end{align}
where $G_0=\vert g_0\vert^2$ is the peak FEL gain, and $R=\vert
r\vert^2$ is the peak reflectivity of the feedback system.
Regenerative amplification requires that $G_{\text{eff}}
>1$. Note that $G_{\text{eff}}$
depends on the time-bandwidth product $\sigma_m\sigma_\tau$, but
not on $\sigma_m$ or $\sigma_\tau$ separately.

The filtered radiation power at the undulator entrance for $n\gg
1$ is then
    \be
    P_n(t) =\vert E_n\vert^2 = P_0 G_{\text{eff}}^n \exp\left(-{t^2\over 2\sigma_{x0}^2}\right) \,,
    \ee
where $P_0$ is the effective noise power within the narrow
bandwidth that starts the process. The amplified radiation at the
end of the $n^{th}$ undulator pass is
    \be
    P_n^a (t) =\vert E_n^a \vert^2 = P_0 G_0 G_{\text{eff}}^n \exp\left(-{t^2\over 2 \sigma_{xa}^2}\right)
    \,, \label{eq:all rad}
    \ee
with $\sigma_{xa}$ given by Eq.~(\ref{eq:length2}). If we neglect
any absorption in the crystal, the part of the radiation energy
(with frequency content mainly outside the feedback bandwidth) may
be transmitted with the maximum efficiency
    \be
    \eta = {\int P_n^a (t)dt - \int P_{n+1} (t)dt \over \int
    P_n^a(t)dt}= 1 - R \sqrt{4\sigma_m^2\sigma_{xa}^2\over
    1+4\sigma_m^2\sigma_{xa}^2}\,. \label{eq:efficiency}
    \ee
In view of Eq.~(\ref{eq:eff gain}), the maximum extraction
efficiency is also a function of the time-bandwidth product
$\sigma_m\sigma_\tau$.

\begin{figure}
\centering
    \includegraphics[width=0.5\linewidth]{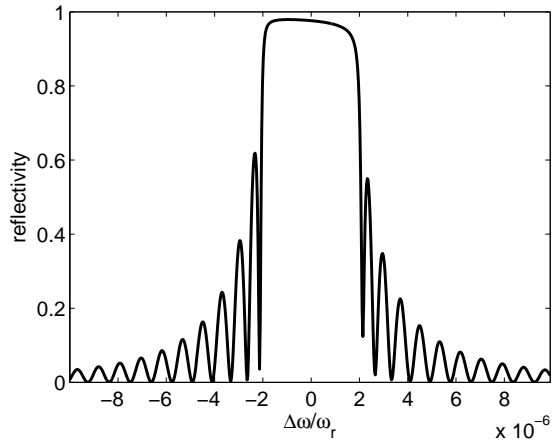}
    \caption{X-ray reflectivity of a 100-$\mu$m-thick diamond (400) crystal for 8-keV, $\pi$-polarized radiation.}
    \label{fig:crystal}
    \vspace{-4mm}
\end{figure}

As a numerical example, we discuss how the proposed RAFEL might be
implemented in the Linac Coherent Light Source
(LCLS)~\cite{LCLS-CDR}. The x-ray wavelength is chosen to be about
1.55~\AA\ since diamond (400) crystals may be used at a Bragg
angle $\theta_B= 60^\circ$. The diamond (115) reflection plane may
be as well chosen at 1.2~\AA\ for the same Bragg angle. Three such
crystals are necessary to form an x-ray cavity as shown in
Fig.~\ref{fig:rafel}. The reflectivity curve of a 100-$\mu$m-thick
diamond (400) crystal for the 8-keV, $\pi$-polarized radiation is
shown in Fig.~\ref{fig:crystal} as computed by XOP~\cite{xop}. The
x-ray reflectivity $R\approx (97\text{\%})^3 \approx 91\text{\%}$
within the Darwin width $\Delta \theta_D \approx 7$~$\mu$rad,
corresponding to the flattop region of Fig.~\ref{fig:crystal} with
$\Delta\omega_m/\omega_r = \Delta\theta_D/\tan\theta_B \approx
4\times 10^{-6}$. The expected rms angular divergence of the FEL
radiation is about 0.5~$\mu$rad, which is well within the Darwin
width but washes out the interference fringes shown in
Fig.~\ref{fig:crystal}. The crystals may be bent slightly to
provide the necessary focusing of the filtered radiation at the
undulator entrance.

\begin{figure}
\centering
    \includegraphics[width=0.5\linewidth]{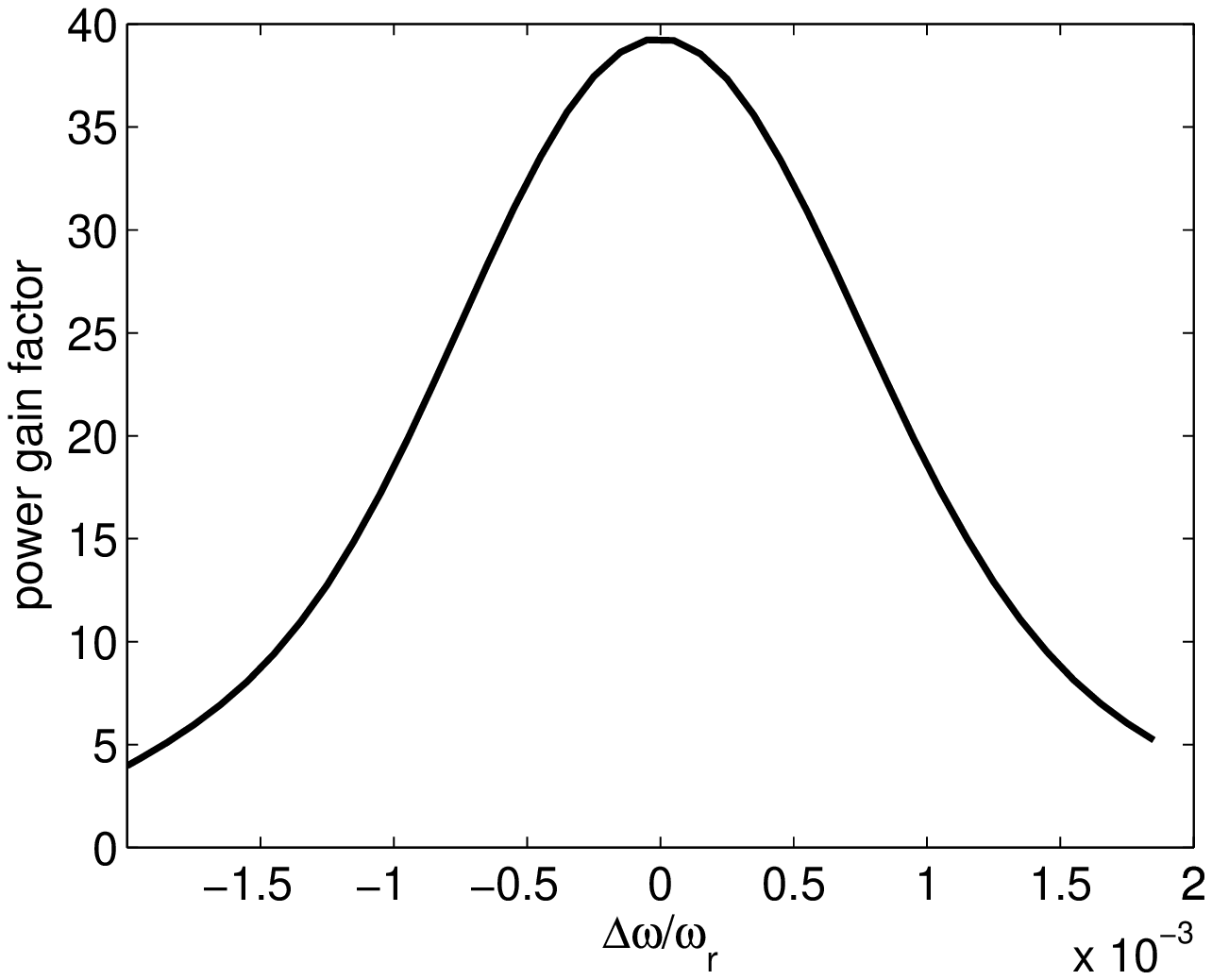}
    \caption{Power gain factor predicted from GENESIS simulation as a function of the relative frequency detune.}
    \label{fig:genesis}
    \vspace{-4mm}
\end{figure}

In order to accelerate a long bunch train in the SLAC linac, we
use the entire rf macropulse available without the rf pulse
compression (SLED). The maximum LCLS linac energy, without the
SLED, is about 10 GeV. Table~\ref{tab:rafel} lists the beam and
undulator parameters that are typical for x-ray FELs such as the
LCLS, except that the length of the undulator is only 20~m instead
of more than 100~m planned for the LCLS. We perform the
three-dimensional (3-D) GENESIS~\cite{genesis} FEL simulation that
shows the maximum power gain $G_0\approx 39$ after the 20-m
undulator, with the fwhm relative gain bandwidth about $2\times
10^{-3}$~(see Fig.~\ref{fig:genesis}). The LCLS accelerator and
bunch compressor systems are expected to generate a bunch current
profile which is more flattop than Gaussian, with a flattop
duration $T=100$~fs~\cite{LCLS-CDR}. If we take
$\sigma_\tau\approx T/2.35$ and $\sigma_m\approx
\Delta\omega_m/2.35$ in Eq.~(\ref{eq:eff gain}), we obtain the
round-trip gain $G_{\text{eff}}\approx 16$ under these parameters.

\begin{table}
\begin{center}
\caption{\label{tab:rafel} Parameters for an x-ray RAFEL.}
\begin{tabular}{|llc|}
\hline Parameter & Symbol & Value \\
\hline electron energy & $\gamma mc^2$ & 9.9 GeV
\\ number of bunches & & 10 to 11
\\ bunch spacing & & $\sim 0.25$ $\mu$s
\\ bunch charge & $Q$ & $\sim 300$ pC
\\ bunch peak current & $I_{pk}$ & 3 kA
\\ fwhm bunch duration (flattop) & $T$ & 100 fs
\\ rms energy spread at undulator & $\sigma_E/E$ & $1\times 10^{-4}$
\\ transverse norm. emittance & $\gamma\varepsilon_{x,y}$ & 1 $\mu$m
\\ undulator mean beta function & $\beta_{x,y}$ & 18 m
\\ undulator period & $\lambda_u$ & 0.03 m
\\ undulator parameter & $K$ & 2.4
\\ FEL wavelength & $\lambda_r$ & 1.55~\AA
\\ photon energy & $\hbar\omega_r$ & 8~keV
\\ FEL parameter & $\rho$ & $5\times 10^{-4}$
\\ undulator length & $L_u$ & $20$~m
\\ maximum FEL gain per pass & $G_0$ & $39$
\\ 3-crystal bandwidth  & $(\Delta\omega_m/\omega_r)$ & $4\times 10^{-6}$
\\ 3-crystal reflectivity & $R$ & $91\%$
\\ \hline
\end{tabular}
\end{center}
    \vspace{-4mm}
\end{table}

We have developed a 1-D FEL code that simulates the regenerative
amplification process. The electron rms energy spread is increased
in the 1-D code to $3.8\times 10^{-4}$ so that the 1-D FEL gain
matches the 3-D FEL gain $G_0=39$ determined by parameters in
Table~\ref{tab:rafel}. The simulation using a flattop current
profile and a nearly flattop crystal reflectivity curve shows that
the round-trip gain $G_{\text{eff}}\approx 14$ in the exponential
growth stage and that the RAFEL reaches saturation within 10 x-ray
passes. For a total x-ray cavity length of 75 m (25 m for each of
three cavity arms in Fig.~\ref{fig:rafel}), the duration of the
10-bunch train is about 2.25~$\mu$s, well within the 3.5-$\mu$s
uncompressed rf pulse length even after taking into account the
structure filling time ($\sim 0.8$~$\mu$s). The beam loading is
estimated to be small for less than 2 mA average current within
the bunch train. To stay within the FEL gain bandwidth as shown in
Fig.~\ref{fig:genesis}, the relative energy variation within the
bunch train should be less than $\pm$0.05\%. A bunch-to-bunch time
jitter of about $\pm$100~fs would require a 11-bunch train of
2.5~$\mu$s in order for the FEL to reach saturation.

\begin{figure}
\centering
    \includegraphics[width=0.5\linewidth]{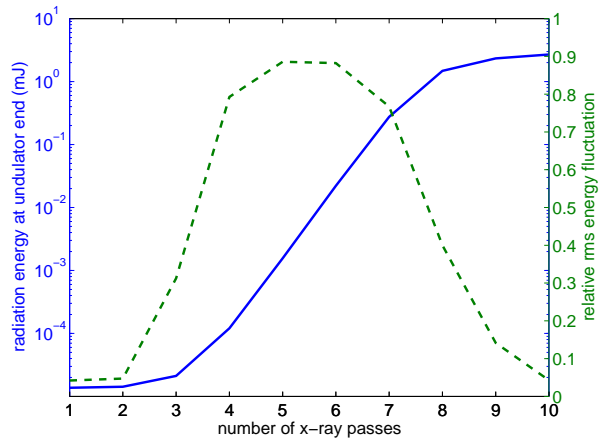}
    \caption{(Color) Average radiated energy (blue solid line) and relative rms energy fluctuation (green dashed line) at the undulator end.}
    \label{fig:RadEnergy}
    \vspace{-4mm}
\end{figure}

\begin{figure}
\centering
    \includegraphics[width=0.5\linewidth]{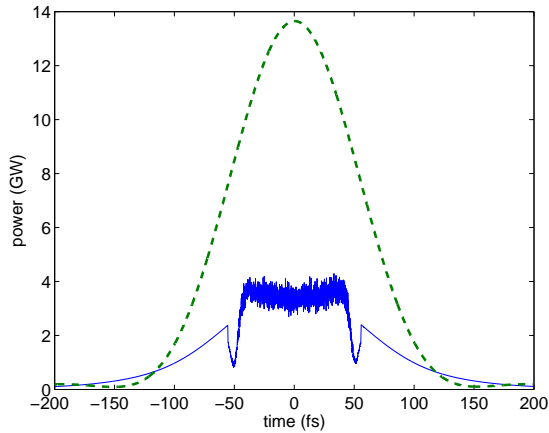}
    \caption{(Color) Temporal profile of the reflected (green dashed line) and
    transmitted (blue solid line) FEL power at the end of 10$^{th}$ pass.}
    \label{fig:FilteredPower}
    \vspace{-4mm}
\end{figure}

\begin{figure}[t]
\centering
    \includegraphics[width=0.5\linewidth]{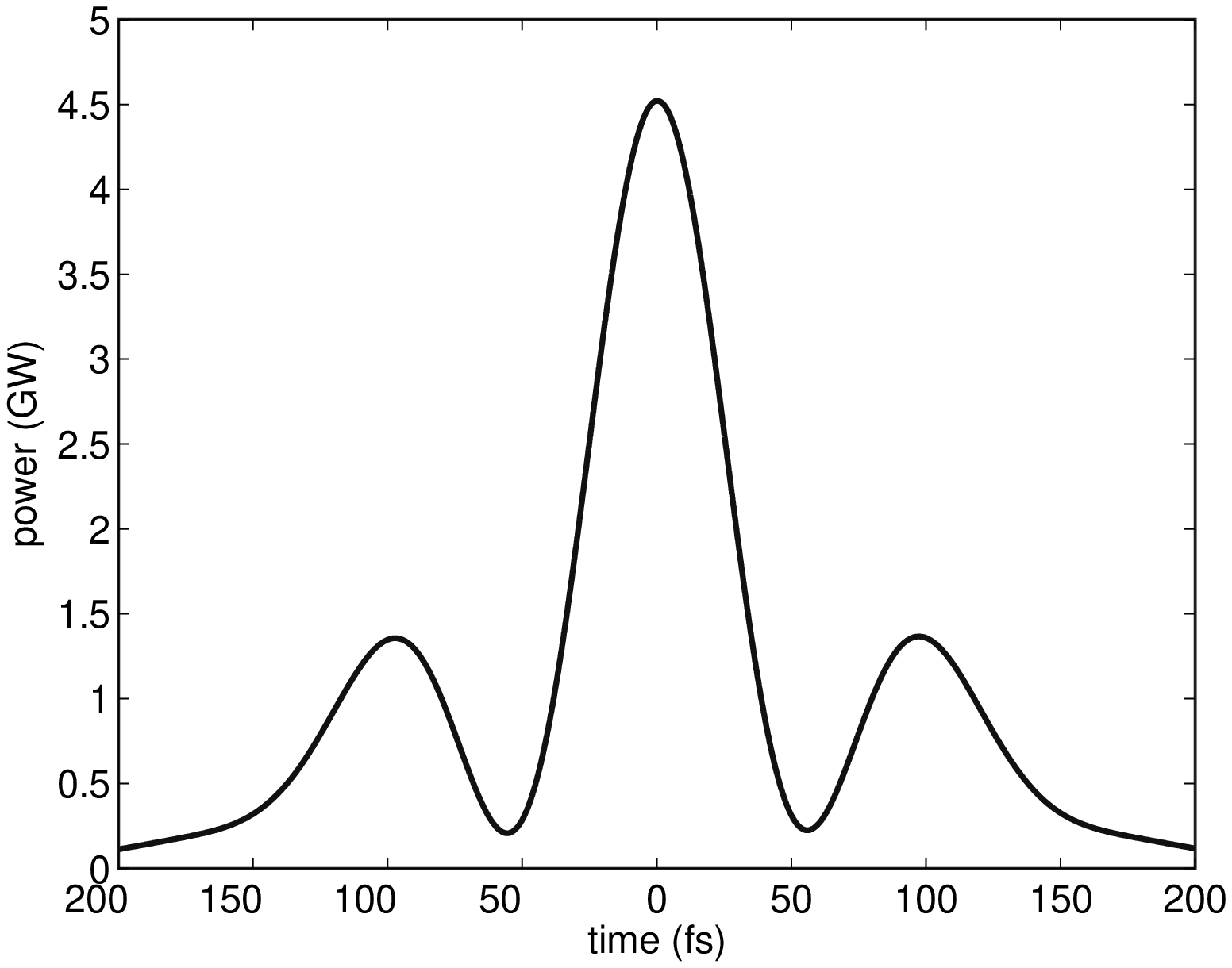}
    \caption{Temporal profile of the final transmitted FEL power after
    passing a monochromator with a fwhm bandwidth $2\Delta\omega_m/\omega_r=8\times 10^{-6}$ to filter out the SASE radiation.}
    \label{fig:TransmittedPower}
   \vspace{-4mm}
\end{figure}

Figure~\ref{fig:RadEnergy} shows that the radiation energy at the
undulator end is mainly the broadband SASE radiation in the first
three passes or so and is then dominated by the narrow bandwidth
filtered signal up to the FEL saturation.
Figures~\ref{fig:FilteredPower} shows the temporal profile of the
reflected and transmitted FEL power for a 100-$\mu$m-thick diamond
crystal with about 82\% transmission outside the crystal bandwidth
around 8 keV. The broadband SASE radiation transmitted through the
end crystal (the noisy part of the blue solid curve in
Fig.~\ref{fig:FilteredPower}) can be separated from the
narrow-bandwidth signal by another monochromator following the
transmission as demonstrated in Fig~\ref{fig:TransmittedPower}.
The total x-ray energy dose absorbed by the undulator-end crystal
(FEL plus spontaneous radiation) is estimated to be two orders of
magnitude smaller than the melting dose level for diamond.
Finally, Fig.~\ref{fig:RadEnergy} also shows that the shot-to-shot
radiation energy fluctuates up 90\% in the exponential growth
stage but quickly reduces to about 5\% at the end of the 10$^{th}$
pass. Although a monochromator may also be used in a saturated
SASE FEL to select a single longitudinal mode, the radiation power
will be reduced by the ratio of the SASE bandwidth to the
monochromator bandwidth, and the filtered radiation energy still
fluctuates 100\%.

While we consider a ring x-ray cavity with $60^\circ$ Bragg
reflection for illustration, the RAFEL scheme and its analysis
presented in the paper is equally applicable to a backscattered
x-ray cavity with $90^\circ$ Bragg reflection. The round-trip time
of such a cavity is only two thirds of the ring cavity shown in
Fig.~\ref{fig:rafel}, allowing for 50\% more electron bunches in a
bunch train of the same duration to participate in the RAFEL
process. The reflectivity at exactly $90^\circ$ Bragg reflection
for cubic crystals such as diamond may be complicated by
multiple-wave diffraction and has not been studied here. Crystals
with lower structure symmetry such as sapphire may provide the
necessary high reflectivity in backscattering as demonstrated in
Ref.~\cite{shvydko}.

In summary, we have described a narrow-bandwidth regenerative
amplifier FEL (RAFEL) at the hard x-ray wavelength region using
Bragg crystals that produces nearly transform limited x-ray pulses
in both transverse and longitudinal dimensions. Compared to a SASE
x-ray source that possesses a typical bandwidth on the order of
$10^{-3}$, the bandwidth of an x-ray RAFEL can be more than two
orders of magnitude smaller, resulting in a factor of a few
hundred improvement in spectral brightness of the radiation
source. The use of multiple bunches in a bunch train for
regenerative amplification allows for a relatively short undulator
system and may be adapted in the LCLS using the SLAC s-band linac.
Since superconducting rf structures can support a much longer
bunch train in an rf macropulse, an x-ray RAFEL based on a
superconducting linac may require a much lower single pass gain
and hence relax some of beam and jitter requirements provided that
the additional radiation damage to the x-ray optics is tolerable.
Therefore, the method described in this paper is a promising
approach to achieve a fully coherent x-ray laser.

We thank J. Hastings and J. Arthur for useful discussions on x-ray
optics. Z. H. thanks K.-J. Kim for general discussions and for
providing Refs.~\cite{colella,adams97}. This work was supported by
Department of Energy contracts DE--AC02--76SF00515.


\end{document}